\documentclass{blux}
\usepackage{mathptmx}
\usepackage{amsfonts}
\usepackage{amsmath}
\usepackage{mleftright}
\usepackage{arydshln}
\usepackage{euscript}
\usepackage{graphicx}
\usepackage{breqn}

\usepackage[scaled=0.92]{helvet}
\usepackage{courier}
\title{On discretization of continuous-time LPV control solutions}
\author{Yorick Broens$^1$, Hans Butler$^{1,2}$ and Roland T\'oth$^{1,3}$ \\
       $^1$ Control Systems Group, Eindhoven University of Technology, \ Eindhoven, The Netherlands \\ 
       $^2$ ASML, Veldhoven, The Netherlands \\ 
       $^3$ Systems and Control Laboratory, Institute for Computer Science and Control, Budapest, Hungary\\
    Email: {\tt Y.L.C.Broens@tue.nl} 
    }
\date{%
$^1$ Control Systems Group, Eindhoven University of Technology\\
$^2$
}

\begin{document}
\maketitle
\section{Introduction}
\vspace*{-2.5mm}
In recent years, the Linear Parameter-Varying (LPV) framework has become increasingly useful for analysis and control of time-varying systems. Generally, LPV control synthesis is performed in the continuous-time (CT) domain
due to significantly more intuitive performance shaping methods in CT, see \cite{zhou1998essentials}. However, the main complication of CT synthesis approaches is the successive implementation of the resulting CT control solutions on physical hardware. In the literature, several discretization methods have been developed for LPV systems, see \cite{Tothbook}. However, most of these approaches necessitate heavy nonlinear operations introduced by the discretization of these time-varying matrices or can introduce significant approximation error, thereby severely limiting implementation capabilities of CT LPV control solutions. Alternatively, the $w'$ discretization approach has been introduced in the LTI case to allow for preservation of the CT control, see \cite{whitbeck1978digital}. Based on these observations, this paper aims at extending the $w'$ discretization approach to LPV systems, such that implementation of CT LPV control solutions on physical hardware is simplified.
\vspace*{-4mm}
\section{Approach}
\vspace*{-2.5mm}
Consider a CT LPV state-space (SS) representation, denoted by $G$, which is given by:
\vspace*{-2mm}
\begin{small}
\begin{equation}
\begin{split}
\left[\begin{array}{c}
\dot{x}(t) \\ y(t)
\end{array} \right] = \left[\begin{array}{cc}
A(p(t)) &B(p(t))  \\ 
C(p(t)) &D(p(t)) 
\end{array} \right] \left[\begin{array}{c}
x(t) \\ u(t)
\end{array} \right],
\end{split}
    \label{DTSYSTEM}
\end{equation}
\end{small}

\vspace*{-4mm}
where $p:\mathbb{R}\rightarrow \mathbb{P}\subseteq \mathbb{R}^{n_p}$ corresponds to the scheduling vector, $x:\mathbb{R}\rightarrow \mathbb{X}\subseteq \mathbb{R}^{n_x}$ is the state variable, $u:\mathbb{R}\rightarrow \mathbb{U}\subseteq \mathbb{R}^{n_u}$ denotes the control input and $y:\mathbb{R}\rightarrow \mathbb{Y}\subseteq \mathbb{R}^{n_y}$ corresponds to the output signal. In order to transform the CT LPV SS representation (\ref{DTSYSTEM}) to the $w'$ domain, the frequency domain filter:
\vspace*{-2mm}
\begin{small}
\begin{equation}
    w'= \frac{2}{T_s} \frac{z-1}{z+1},
\end{equation}
\end{small}

\vspace*{-4mm}
is reformulated into an equivalent time-domain operator $r$:
\vspace*{-2mm}
\begin{small}
\begin{equation}
    r = \frac{2}{T_s} \frac{q-1}{q+1},
\end{equation}
\end{small}

\vspace*{-5mm}
where $q$ corresponds to the shift operator and $T_s$ is the sampling time. Moreover, the system represented by ($\ref{DTSYSTEM}$) is expressed in the $w'$ domain as:
\vspace*{-2mm}
\begin{small}
\begin{equation}
\begin{split}
\left[\begin{array}{c}
rx(k) \\ y(k)
\end{array} \right] = \left[\begin{array}{cc}
A(p(k)) &B(p(k))  \\ 
C(p(k)) &D(p(k)) 
\end{array} \right] \left[\begin{array}{c}
x(k) \\ u(k)
\end{array} \right],
\end{split}
    \label{DTSYSTEM2}
\end{equation}
\end{small}

\begin{figure}[h]
    \centering
    \includegraphics[trim={0.5cm 0cm 0.5cm 0cm},clip,width=\linewidth]{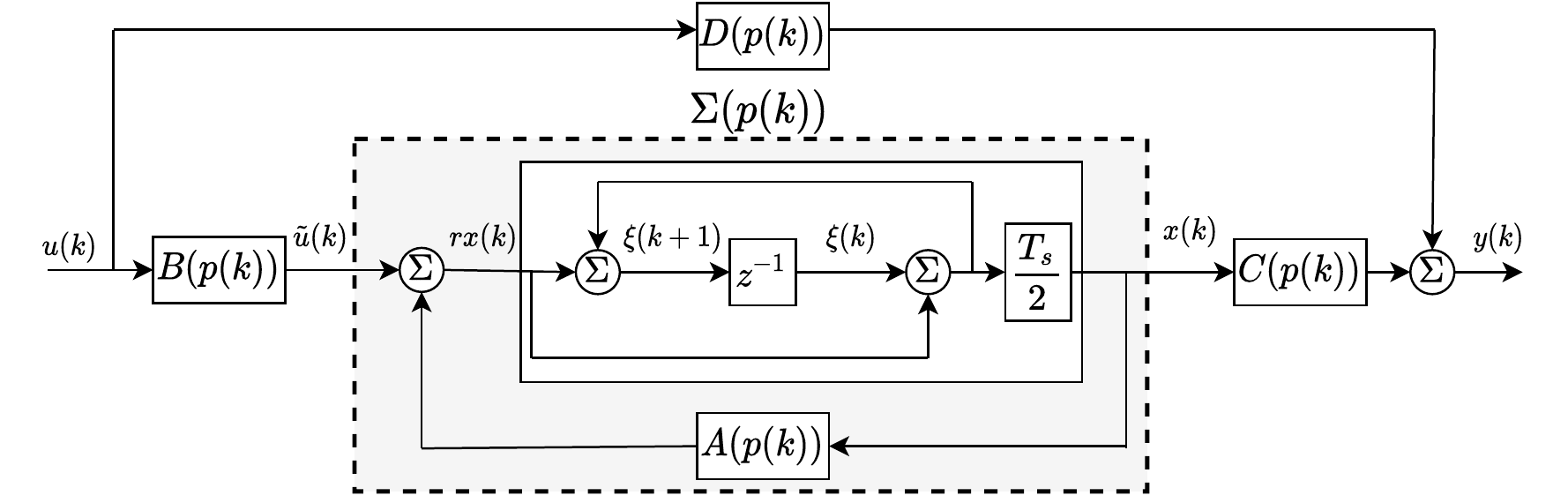}
     \vspace*{-8mm}
    \caption{Block interconnection of the $w'$ implementation, where the white area denotes the $r^{-1}$ operator and $\Sigma(p(k))$ corresponds to the equivalent dynamics, which is used to remove the algebraic loop that is introduced by the $r^{-1}$ operator.}
    \label{fig:my_label}
\end{figure}

\vspace*{-5mm}
where the CT LPV SS matrices are preserved by defining the $r^{-1}$ operator, see Figure \ref{fig:my_label}, which corresponds to:
\vspace*{-2mm}
\begin{small}
\begin{equation}
    \begin{split}
\left[\begin{array}{c}
\xi(k+1) \\ x(k)
\end{array} \right] = \left[\begin{array}{cc}
I&2I\\ \frac{T_s}{2}I &\frac{T_s}{2}I 
\end{array} \right] \left[\begin{array}{c}
\xi(k) \\ rx(k)
\end{array} \right]
\end{split}
\end{equation}
\end{small}

\vspace*{-5mm}
From the structure of the $r^{-1}$ operator, it is observed that it is identical to the Tustin operator, see \cite{Tothbook}. Nonetheless, the proposed approach allows for discrete-time implementation using the CT LPV system matrices, thereby both preserving physical insight and simplification of the implementation procedure of CT LPV control solutions on physical hardware.
In order to remove the algebraic loop that is introduced by the $r^{-1}$ operator, see Figure \ref{fig:my_label}, subsystem $\Sigma(p(k))$ is introduced, which is given by: 

\vspace*{-8.5mm}
\begin{small}
\begin{equation}
    \begin{bmatrix}\xi(k+1) \\ x(k) \end{bmatrix} = 
    \begin{bmatrix}
    I+\Phi(p(k))A(p(k))T_s & 2\Phi(p(k)) \\
    \Phi(p(k))\frac{T_s}{2} & \Phi(p(k))\frac{T_s}{2}
    \end{bmatrix} \begin{bmatrix}
    \xi(k) \\ \Bar{u}(k)
    \end{bmatrix},
\end{equation}
\end{small}

\vspace*{-5.5mm}
where \begin{small}$\Phi(p(k)) = [I-A(p(k))\frac{T_s}{2}]^{-1}$\end{small}. Moreover, the full discretized system, denoted by $G_d$, is given by:
\vspace*{-2mm}
\begin{small}
\begin{equation}
    G_d = C(p(k))\cdot \Sigma(p(k)) \cdot B(p(k)) + D(p(k)),
\end{equation}
\end{small}
 
\vspace*{-8mm}
where \begin{small}$\text{det}\left(I-A(p(k))\frac{T_s}{2}\right) \neq 0 \ \forall \  p \in \mathbb{P}$\end{small}.
\vspace*{-4mm}
\section{Conclusions}
\vspace*{-3mm}

It is observed that application of the $w'$ discretization is equivalent with the Tustin discretization using the CT LPV SS matrices. Furthermore, implementing the $w'$ discretization via the $r^{-1}$ operator simplifies the implementation process of CT LPV control solutions on physical hardware while the physical interpretation of the CT LPV control solution is preserved.
\vspace*{-6.5mm}
\bibliographystyle{ieeetr}        % Include this if you use bibtex 
\bibliography{MyBib}              % and a bib file to

\end{document}